\documentclass[a4paper,10pt]{article}
\usepackage{enumerate}
\usepackage{color}
\usepackage[utf8]{inputenc} 
\usepackage[english]{babel}
\usepackage[T1]{fontenc}
\usepackage{graphicx}
\usepackage{amsfonts,amssymb,amsmath,latexsym,amsthm}
\usepackage{textcomp}
\usepackage[pdftex]{hyperref}
\usepackage{geometry}
\usepackage[toc,page]{appendix}
\DeclareMathOperator{\sech}{sech}

\begin{document} 

\title{On the linearity and stability of electrostatic structures based on the Schamel equation}
\author{Hans Schamel$^{1}$, Efim Pelinovsky$^{2,3}$ and Marcelo V. Flamarion$^{4}$}
\date{}

\maketitle
\begin{center}
{\footnotesize $^1$Physikalisches Institut, Universit\"at Bayreuth, 95440 Bayreuth, Germany}  \\

\vspace{0.3cm}
{\footnotesize 
$^{2}$Faculty of Informatics, Mathematics and Computer Science, HSE University, Nizhny Novgorod 603155, Russia.

$^{3}$ Gaponov-Grekhov Institute of Applied Physics, Nizhny Novgorod, 603122, Russia. \\
$^{4}$ Departamento Ciencias—Secci{\' o}n Matem{\' a}ticas, Pontificia Universidad Cat{\' o}lica del Per{\' u}, Av. Universitaria 1801, San Miguel 15088, Lima, Peru.}




\end{center}


\begin{abstract} 
This paper contributes in the first part to the correct understanding of the linear limit in the Schamel equation (S-equation) from the perspective of structure formation in collisionless plasmas.{ The corresponding modes near equilibrium turn out to be nonlinear modes of the underlying microscopic Vlasov-Poisson (VP) system,  for which particle trapping is essential and which propagate with one of the  slow acoustic velocities.} A simple shift of the electrostatic potential to a new pedestal leads to non-negativity and thus mitigates the positivity problem of the S-equation. The stability of a solitary electron hole (bright soliton), based on both the S-equation and an earlier transverse but limited VP instability analysis, exhibits marginal longitudinal stability and linear perturbations in the form of the asymmetric shift eigenmode of a solvable Schrödinger problem. {This finding on the predominance of shift-mode perturbation thus provides a new clue for the general kinetic proof of the marginal stability and transversal instability of electrostatic structures including undisclosed potentials.}
	\end{abstract}

\section{Introduction}
  The Schamel equation was originally formulated to investigate undamped electrostatic waves in a Maxwellian plasma environment \cite{Schamel:1972}. Schamel's work focused on the complex interactions between ions and electrons, incorporating aspects such as plasma density {and temperature variations, particle trapping and associated scenarios, non-Landau behavior, negative energy holes and attractors, cusp-singularities, collisions and anomalous resistivity, holes in synchrotrons, to mention a few.  } Since its inception, the Schamel equation has become an essential framework for exploring ion acoustic wave dynamics in plasma systems \cite{Schamel:1973, Ali:2017, Chowdhury:2018, Mushtaq:2006, Williams:2014, Saha:2015a, Saha:2015b} and recently in metamaterials \cite{Mogilevich:2023, Zemlyanukhin:2019}, damping systems \cite{Shan:2019, Sultana:2022} and electrical circuits \cite{Asif:2020, Kengne:2020}.
 
The Schamel equation is frequently compared to the well-known Korteweg–de Vries (KdV) equation \cite{Zabusky:1965}, as both share the same dispersion relation. The original form of the Schamel equation lacks a modulus in its nonlinear term, meaning it can be expressed in a canonical form as 
 \begin{equation}\label{Sch1}
 \phi_t+\sqrt{\phi}\phi_x+\phi_{xxx}=0.
 \end{equation}
However, to address practical considerations, particularly in numerical studies, mathematicians have introduced a modified version known as the modular Schamel equation
 \begin{equation}
 \phi_t+\sqrt{|\phi|}\phi_x+\phi_{xxx}=0.
 \end{equation}
This modular form is more suited for numerical purposes. It is often compared to the modified KdV (mKdV) equation, as both equations support solitary wave solutions of either polarity, a key feature in the study of nonlinear wave dynamics. From a mathematical standpoint, the Schamel equation presents unique challenges due to its square-rooted nonlinear term, and in some cases, the modulus nonlinear term. These features complicate both analytical and numerical treatments. Additionally, unlike integrable equations such as the Gardner equation or the modified Korteweg–de Vries (mKdV) equation, the Schamel equation lacks integrability. This non-integrability introduces further complexity when attempting to derive exact solutions or apply standard analytical techniques commonly used in plasma physics studies \cite{Ruderman:2008, Ruderman:2023, Schamel:1972, Schamel:1973}. Recent studies by Flamarion et al. \cite{Flamarion:2023a} and Didenkulova et al. \cite{Flamarion:2023b} have explored the interaction of solitary wave solutions in the context of the Schamel equation. Although the Schamel equation is nonintegrable, the collisions between solitary waves are nearly elastic, meaning that after interaction, the solitary waves largely retain their original form. However, Didenkulova et al. \cite{Flamarion:2023b} noted that in bipolar interactions, energy tends to transfer from the smaller wave (in modulus) to the larger one. This energy redistribution, combined with the dispersive tails generated during the collision, contributes to the formation of freak waves. A more detailed explanation of this freak wave formation mechanism was provided in a subsequent work \cite{Flamarion:2024}.
 
The response of solitary waves to external forces has also been a focus of study within the framework of the Schamel equation. For instance, Chowdhury et al. \cite{Chowdhury:2018} derived a forced version of the Schamel equation, incorporating the effects of an external time-dependent force. In a related study, Flamarion and Pelinovsky \cite{Flamarion:2023} explored the phenomenon of trapped waves, a well-known concept in nonlinear physics. In this scenario, a solitary wave resonates with the external force, exhibiting behavior akin to that of nonlinear oscillators in classical mechanics. A damped Schamel equation, in canonical form
 \begin{equation}
 \phi_t+\sqrt{\phi}\phi_x+\phi_{xxx}+C\phi=0.
 \end{equation}
where, $C$ is a constant has been introduced in the literature \cite{Shan:2019}. Shan \cite{Shan:2019} investigated the nonlinear behavior of high-frequency electron-acoustic (EA) waves in a dissipative plasma, which consists of a cold beam electron fluid, Schamel-kappa distributed hot trapped electrons, and stationary ions. By applying the multiple scale expansion method, Shan derived this damped Schamel equation to model small-amplitude electrostatic potential disturbances while accounting for dissipative effects. Sultana and Kourakis \cite{Sultana:2022} later employed this equation in their analysis of electrostatic potential, examining the nonlinear characteristics of dissipative ion-acoustic solitary waves in the presence of trapped electrons.

In recent decades it has become clear that the equilibrium theory presented in \cite{Schamel:2023,SChakrabarti:2023} is the appropriate, indeed the only method to completely describe the spectrum of long-lived electrostatic structures in VP plasmas. The reason for this is that only mathematically sound distribution functions are used and a veritable treasure trove of free parameters is available for structural adjustment.
 In this theory, the t-independent Vlasov equations for electrons and ions are solved exactly using the two concepts of constants of motion and trapping scenarios (TS), respectively. In this method, also known as the Schamel method, self-consistency is achieved by using the pseudo-potential.
 Under certain circumstances, however, a simpler description is available, namely when the  structures move at one of the three acoustic speeds, which are the slow ion acoustic (SIA), the ion acoustic (IA) and the slow electron acoustic (SEA) speed. One can then resort to S-equations that are formulated in macroscopic space-time $(x,t)$ instead of in microscopic phase space $(x,v,t)$ for electrons (or $(x,u,t)$ for ions) and are thus predestined to describe the structural dynamics on a much simpler basis. This reduction reflects the mathematical realization of a strong convergence in macroscopic dynamics, while in microscopic dynamics only a weak convergence is found \cite{Mouhout:2011}.

The present paper addresses, among other things, the question of what the necessary non-negativity of the solution to an S-equation means and how misunderstandings or misinterpretations can be avoided. \\
As explained in more detail in the review article \cite{Schamel:2023}, the S-equation is an asymptotic evolution equation. In the case of a collisionless plasma, it describes the spatio-temporal behavior of electrostatic structures $\phi(x,t)$ after the violent particle trapping processes have already taken place. This means it only takes into account plasmas states when they have entered into the "calmer waters", just before reaching the stationary equilibrium state. It is therefore not suited as a model that includes linear waves from fluid or linearized Vlasov descriptions because trapping effects are neglected in these approaches. It also fails of course to solve Cauchy's initial value problem of the Vlasov-Poisson system (VP system) in which $\{f_e(x,v,0),f_i(x,u,0),\phi(x,0)\}$ are prescribed as initial values. The latter is a mathematically intractable problem because stochasticity and non-integrability play a role during the trapping process; these are processes such as folding, trapping, detrapping, filamentation, etc., which occur in resonant wave-particle interaction in phase space resulting in a non-treatability.\\
This means that all references to linear wave solutions, whether from the fluid description or the linear Vlasov description, can be forgotten. Our solutions have already passed through and left this early stage of evolution. However, linearization within the S-equation is still possible, but must be understood in the following sense. 

We return to the theory of equilibrium structures in VP systems \cite{Schamel:2023} and consider a positive potential structure, which propagates in the laboratory frame with the velocity $v_0$. It is based on the positive pedestal:
\[
0 \le \phi(x - v_0 t) \leq \psi <<1
\]
and it is assumed that it is generated exclusively by an electron trapping scenario  in its smoothest form. For simplicity, other trapping scenarios (TS) { inclusively ion trapping effects} are thus neglected. In this case, the governing equations, { whose derivation is repeated in a compact form for the interested reader in the Appendix (where we actually also take into account the ion trapping effect $B_i$)}, read in its simplest form: $v_D=0$, $B_i=0, ...$
\begin{equation} \label{eq1}
k_{0}^2-\frac{1}{2}Z_{r}'\Big(\frac{v_{0}}{\sqrt{2}}\Big)-\frac{\theta}{2}Z_{r}'\Big(\frac{u_{0}}{\sqrt{2}}\Big)=B_e
\end{equation}
and
\begin{equation} \label{eq2}
-\mathcal V(\phi)=\frac{k_0^2}{2}\phi(\psi-\phi)+\frac{B_e}{2}\phi^{2}\Big(1-\sqrt{\frac{\phi}{\psi}}\Big) ,
\end{equation}
where the first part refers to the nonlinear dispersion relation (NDR) that determines $v_0$. The second part is the pseudo-potential that governs the structure $\phi(x)$ itself.
\begin{figure}[h!]
	\centering	
	\includegraphics[scale =1]{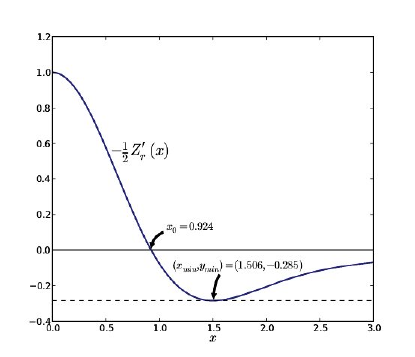}
	\caption{The quantity -$\frac{1}{2}Z_r'$ as a function of $x$, where $Z(z)$ is the complex plasma dispersion function.}
	\label{Fig1}
\end{figure}

{ Figure \ref{Fig1} shows $-\frac{1}{2}Z_r'(x)$ as a function of $x$. Its zero point at $x=0.924$ defines the slow acoustic phase velocities via the nonlinear dispersion relation, namely the electronic with $v_0= 0.924 \sqrt2=1.307$ and the ionic with $u_0:=\sqrt{\frac{\theta}{\delta}}v_0=1.307$. Both corresponding mode structures can be macroscopically described by a Schamel equation.
\begin{figure}[h!]
	\centering	
	\includegraphics[scale =0.34]{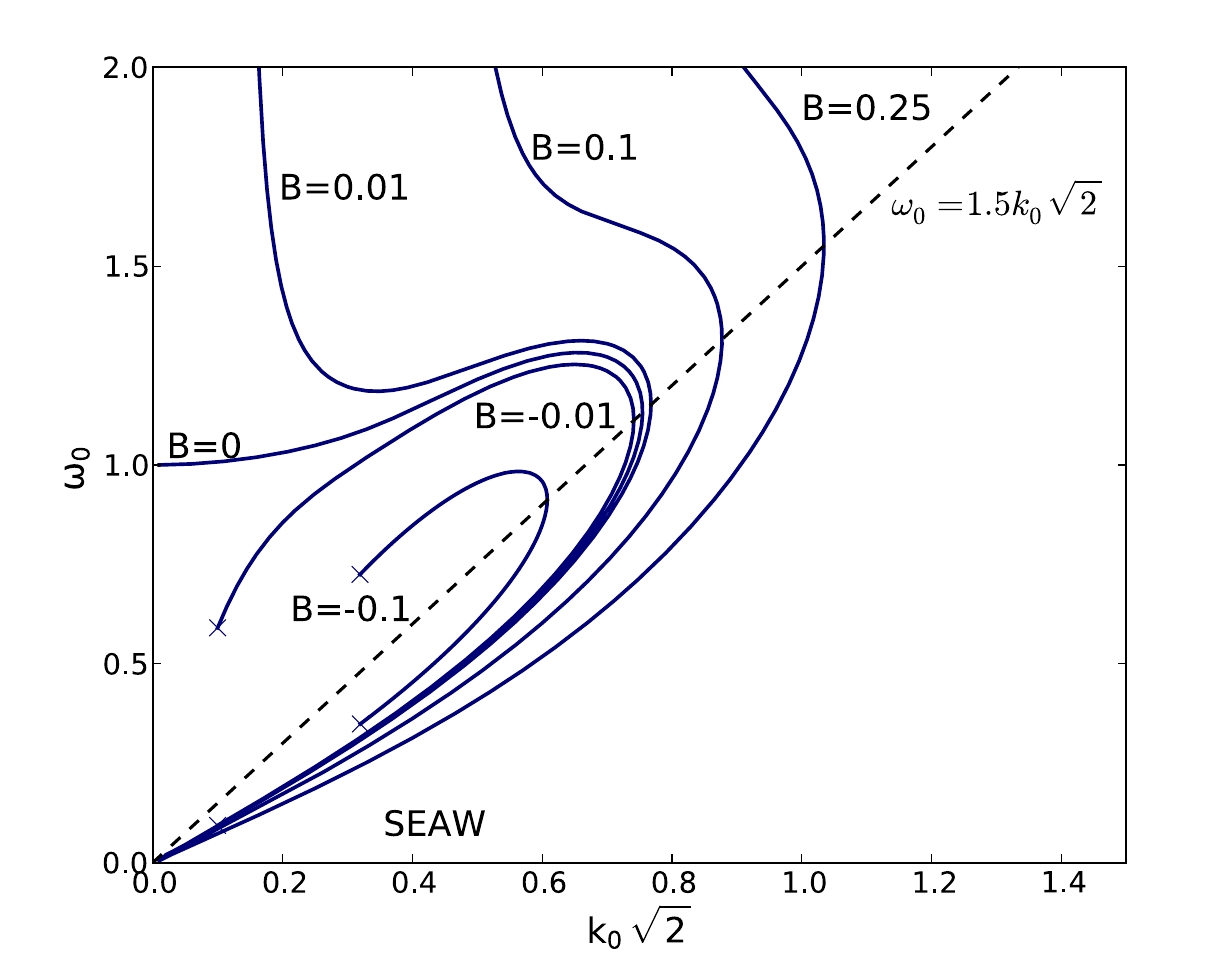}
	\caption{The high-frequency electronic part of the NDR (4) with $B:=B_e$}
	\label{Fig2}
\end{figure}

Figure \ref{Fig2} shows the high-frequency electronic part of the NDR (4) with $B:=B_e$, indicating a multitude of new solutions besides the Langmuir and the slow electron-acoustic branch (given by $B=0, k_0<<1$).

}

 { The pseudo-potential (\ref{eq2}) in its canonical form} has the noteworthy property that it is independent of the phase velocity $v_0$ and thus cannot be used for its determination. This system represents cnoidal waves and has been thoroughly discussed as a two parametric system ($k_0, B_e$) by Korn \& Schamel \cite{Korn:1996}. The potential structure, with the exception of the two solitary parts, is periodic in space and is characterized by the parameter $S = 4B_e/ k_0^2$ which lies in the interval $-8 \leq S \leq \infty$ i.e. between the bounds $S = -8$ and $S = \infty$. It is hence embedded between a solitary hump ("bright soliton"), when $k_0 = 0$, and a solitary dip ("dark soliton"), when $k_0^2 = -B_e/2>0$. The $k_0 = 0$ solitary wave is the well-known:
\begin{equation} \label{eq3}
\phi(x) = \psi\sech^{4}\left( \frac{\sqrt{B_e}}{4}x \right)
\end{equation}
whereas the solitary dip solution, shown in Fig.3 of \cite{Schamel:2023} or Fig.1 of \cite{SChakrabarti:2023}, is more complicated. \\
Of special interest is thereby the single harmonic wave solution which lies  in between at S=0 and is given by:
\begin{equation} \label{eq4}
\phi(x) = \frac{\psi}{2} (1 + \cos k_0 x) \geq 0
\end{equation}
It refers to $B_e = 0$ and is the linearized solution within Schamel's theory, since the trapping nonlinearity, represented by the $B_e$ term, vanishes. \\
This linear wave solution, which is lifted by $\psi/2$ in comparison with an ordinary sinusoidal wave, is therefore strictly  non-negative as all structures are. It has, as said, nothing to do with the ordinary linear waves stemming from linearized Vlasov-Poisson system (or fluid system). The reason is that the electron distribuition $f_e(x-v_0t, v)$ is still nonlinearly distorted in the resonant trapped electron range even when $B_e = 0$. 

This applies to the two S-equations, as well, since they represent these equilibria in the special case of the two acoustic phase velocities, namely when $v_0 \approx \sqrt{\delta}$ (the ion acoustic wave limit, IAW) and when $v_0 = 1.307$ (the slow electron acoustic wave limit, SEAW), respectively,  where $\delta$ is the mass ratio $m_e / m_i$. 

\section{The Dark Soliton and the Negative Pedestal}

In a short insert, we point out that sometimes a negative pedestal is useful for non-positive structures
\[
 -1<<-\psi \leq \hat \phi(x - v_0 t)\leq 0.
\] 
 An example is the ion hole, where ion trapping plays the crucial role. This mode rests on the slow ion acoustic velocity, $u_0=1.307$, and the solitary wave potential assumes the form $\hat \phi(x)=-\psi \sech^4 \left( \frac{\sqrt{B_i}}{4}x \right)$, as first developed in \cite{SBujarbarua:1980}. In the present case, it is the dark soliton that could benefit from this shift of $\phi(x)$ to $\hat\phi(x):=\phi(x)-\psi$, provided that the pseudo-potential and hence the S-equation take a simpler form in this new variable. To check this we first note that the pseudo-potential in $\phi$ becomes for the dark soliton:
\[
-\mathcal V(\phi)=\frac{B_e}{4} \phi (3\phi - \psi - \frac{2}{\sqrt \psi} \phi^{3/2})
\]
 which translates into
\[
-\mathcal V(\hat\phi)=\frac{B_e}{2} (\hat\phi+\psi) [\frac{3}{2}\hat\phi + \psi - \frac{1}{\sqrt \psi}(\hat \phi+\psi)^{3/2}]
\]
 written in the new dependent variable $\hat\phi$. However, as it turns out, our hope is not fulfilled.
The complexity of $\phi(x)$ in the case of dark solitons also carries over to the new expression, which has not become much simpler. The complexity remains and is thus confirmed.\\
Note that the equivalence of both pedestals was proven in \cite{Schamel:2000}. Of course, one can also use an intermediate pedestal, such as $-\psi/2 \le \tilde \phi \le \psi/2<<1$ with $\tilde \phi$ as the new dependent variable. However, this does not increase the attractiveness of the equations and formulas.

\section{The Two Acoustic Solutions of the NDR}
In the present limited case of electron trapping only, the NDR has two acoustic solutions, the ion acoustic solution (IA) and the slow electron acoustic solution (SEA), as already mentioned. The former is obtained by using the approximation
\begin{equation} \label{eq5}
-\frac{1}{2}Z_{r}'\Big(\frac{v_{0}}{\sqrt{2}}\Big)\approx 1-v_0^2+\cdots
\end{equation}
\begin{equation} \label{eq6}
-\frac{1}{2}Z_{r}'\Big(\frac{u_{0}}{\sqrt{2}}\Big)\approx-\frac{1}{u_0^2}\Big(1+\frac{3}{u_0^2}+\cdots\Big)
\end{equation}
and results in
\begin{equation}  \label{eq7}
v_0= \sqrt{\delta}(1-K_1), \mbox{ where } K_1=\frac{k_0^2}{2}-\frac{3}{2\theta}-\frac{\delta}{2}-\frac{B_e}{2},
\end{equation}
where $|K_1|$ is assumed to be small. The latter uses
\begin{equation} \label{eq8}
-\frac{1}{2}Z_{r}'\Big(\frac{v_{0}}{\sqrt{2}}\Big)\approx  \frac{1.307-v_0}{1.307}; \;\ -\frac{1}{2}Z_{r}'\Big(\frac{u_{0}}{\sqrt{2}}\Big)\approx  \frac{-\delta}{\theta v_0^2}; \;\  u_{0}=\sqrt{\frac{\theta}{\delta}}v_0 \mbox{ and } \theta=T_e/T_i,
\end{equation}
and yields
\begin{equation}\label{eq9}
v_0=1.307(1+K_2), \mbox{ where } K_2=k_0^2-\frac{\delta}{1.307^2}-B_e.
\end{equation}

\section{{An alternative, generalized access to the Schamel equation}}
The new concept for the derivation of the S-equation, which represents the entire spectrum of wave structures in the two acoustic limits, has been introduced in \cite{Schamel:2020}. 
It consists in the Ansatz
\begin{equation} \label{eq10}
\Big( \phi_t + v_0 \phi_x \Big) + c \Big[ - \mathcal V''(\phi) \phi_x - \phi_{xxx}\Big]= 0 
\end{equation}
and arises from the fact that in equilibrium $\phi(x - v_0 t)$ both expressions $\left( \cdot \right)$ and $\left[ \cdot \right]$ vanish identically. The first is obvious and the second is obtained by the $x$-derivative of Poisson's equation; we hence have added two terms which vanish in equilibrium. Appropriate  space-time dependent solutions should therefore not deviate too much from equilibrium; to what extent, however, remains an open question. \\
To justify this derivation, it can be said that it reproduces the usual results of the reductive perturbation method (RPM) when the latter is feasible, but has a much wider range of applications when, for example, suitable scaling properties are no longer available. This is the case for the majority of structures namely when further trapping scenarios (TS) (not  treated in the present ms) are in action leading to the so-called undisclosed potentials (UP).
Therefore, this derivation here is much broader applicable from the outset.

In equation \eqref{eq10}, $v_0$ has to be taken up to first order, as used in (\eqref{eq7}, \eqref{eq9}, respectively), whereas the coupling constant $c$ depends on the chosen case, becoming $-\sqrt{\delta}/2$ for the IA case and $1.307$ for the SEA case.

\section{The Ion Acoustic Case}
We note that by applying \eqref{eq2}, it holds that
\begin{equation}\label{eq11}
-\mathcal V''( \phi) = -k_0^2 + B_e \left(1 - \frac{15}{8} \sqrt{\frac{\phi}{\psi} }\right)   \end{equation}
such that (\ref{eq10}) becomes
\begin{equation} \label{eq12}
\frac{1}{\sqrt{\delta}}\phi_t+\Big(1+\frac{15}{16}\frac{B_e}{\sqrt{\psi}}\sqrt{\phi} \Big)\phi_x +\frac{1}{2}\phi_{xxx}=0,
\end{equation}
{ which has the form of (\ref{Sch1}) in the frame traveling with ion acoustic velocity ($x-\sqrt \delta t$).}
It is an extended version of the S-equation and applies for the whole spectrum of waves $-8 \leq S \leq \infty$. It thus includes the bright soliton, when $k_0^2 = 0$ or $S = \infty$, and the dark soliton, when $S = -8$, as well as the single harmonic wave, when $S = 0$ or $B_e = 0$. It coincides with the S-equation (15) of \cite{Schamel:2023} for the bright soliton and with (39) of \cite{Schamel:2023} for the dark soliton. \\
There is a difference to (9) in \cite{SChakrabarti:2023} since the $k_0^2$ term is now absent. The reason that in \cite{SChakrabarti:2023} only $v_0 = \sqrt{\delta}$ was used rather than the full expression including $K_1$. In the corrected version, single harmonic waves show dispersion in the ion acoustic limit.

{There is still a slower mode, the slow ion acoustic mode, which propagates with $u_0=1.307$ \cite{Schamel:2023} and which has not been treated in this paper because the ion trapping effect was neglected. However, if we are not mistaken, it is these ion hole-like structures that propagate at the slow ion acoustic speed and not at the usual ion acoustic speed that have been detected in large quantities in the solar wind \cite{Ergun:2024}. These include regions with hot ions $\theta\le 3.5$, which is the existence condition for  ion holes \cite{Chen:2004, Schamel:1983}, as well as regions with Boltzmann electrons ($\beta=1$) or electrons with a flat-topped distribution ($\beta=0$), but also regions where a positive slope of the ion distribution function is no longer necessarily present.}

\section{The slow electron acoustic  case}

Using $v_0$ of (\ref{eq9}), $c = 1.307$ and neglecting the small term proportional to $\delta$, we get from (\ref{eq10})
\begin{equation}\label{eq13}
\frac{1}{1.307}\phi_t+\Big(1-\frac{15}{8}B_{e}\sqrt{\frac{\phi}{\psi}}\Big)\phi_x-\phi_{xxx}=0.
\end{equation}
This slightly modified $S$-equation is again identical to the previous results, such as (5) and (14) of \cite{SChakrabarti:2023} for the bright soliton and (29) of \cite{SChakrabarti:2023} for the dark soliton. In the harmonic case, we get back (\ref{eq9}).

\section{Stability}

The stability of trapped particle equilibria  is within the microscopic VP system of equations a mathematically delicate problem, since it remains unsolved even as a linear stability theory. An exception are single harmonic structures, which turn out to be linearly marginally stable and independent of the drift velocity $v_D$ in a current-carrying plasma \cite{Schamel:2018,Schamel:2023}, a fact that, incidentally, contradicts the linear Landau analysis of linear Vlasov-Poisson plasmas.
There are two main reasons for this failure. First, most authors did not have a clean, internally consistent equilibrium solution $(\phi(x), v_0)$ at their disposal and second, the resulting  eigenvalue problem for the linear perturbation $\phi_1(x,t)$ remained unsolved. Only by a mathematically unproven truncation of the Taylor series of the nonlocal spectral operator of infinite order after the second term could further progress be made. This approach, called fluid limit \cite{Lewis:1979}, was exploited by one of the authors of this study to derive a linear stability theory for solitary structures \cite{Schamel:1982,Schamel:1986,Schamel:2023}, which is exact within its limitation.

It is therefore interesting to see how the achieved marginal stability is also reflected in the S-equation and how its solution can be classified in relation to the overall unknown solution of the problem. If we take the bright soliton solution (\ref{eq3}) as the unperturbed structure $\phi_0(x)$ together with $v_0$ from ({\ref{eq7}) and the S-equation (\ref{eq12}) for the ion acoustic case we get for the first order perturbation $\phi_1(x)$ in linear approximation

\begin{equation} \label{eq14}
\frac{1}{\sqrt \delta}\phi_{1t} +  \phi_{1x} + \frac{15 B_e}{16 \sqrt \psi} \Big (\sqrt{\phi_0} \phi_{1x} +\frac{\phi_{0x}}{2\sqrt{\phi_0}} \phi_1 \Big )+ \frac{1}{2} \phi_{1xxx} = 0 .
\end{equation}
The zeroth order solution for $\phi_0(x-v_0t)$ is satisfied by the soliton solution (\ref{eq3}) itself with $v_0$ from ( \ref{eq7}).
Using the new variables $\tau:=\sqrt \delta t$ , $\xi:=\alpha x$ with $\alpha:=\frac{\sqrt B_e}{4} $ we get the somewhat simpler pde

\begin{equation} \label{eq15}
\phi_{1\tau} + \alpha \phi_{1\xi} + 15 \alpha^3\Big (\sqrt{\frac{\phi_0}{\psi}} \phi_1\Big )_\xi +  \frac{1}{2}\alpha^3 \phi_{1\xi \xi \xi} = 0 .
\end{equation}
Next we show that a steady-state, co-propagating $\phi_1(\xi, \tau) =\varphi\Big(\alpha(x-v_0t)\Big ) =\varphi \Big (\xi -\alpha(1 + 8\alpha^2) \tau \Big )$ is a solution of this equation where we used the simplest nontrivial form of $v_0$ namely $v_0=\sqrt \delta(1+ B_e/2)$. By insertion we find
\begin{equation} \label{eq16}
-8 \varphi_\xi + 15 \Big (\sqrt{\frac{\phi_0}{\psi}} \varphi\Big )_\xi +  \frac{1}{2} \varphi_{\xi \xi \xi} = 0 ,
\end{equation}
which can immediately be integrated, assuming ($\varphi, \varphi_{\xi \xi}$) vanish as $|\xi| \rightarrow \infty$, and get
\begin{equation} \label{eq17}
 \varphi_{\xi\xi} - 16 \varphi +  30 \sech^2(\xi) \varphi = 0 .
\end{equation}
We hence arrived at a  Schr\"odinger eigenvalue problem \cite{Schamel:1982}

\begin{equation} \label{eq18}
\Lambda \eta_n:= \alpha^2 \Big (\partial_\xi ^2 - 16 + 30 \sech^2(\xi) \Big ) \eta_n= -\lambda_n \eta_n,
\end{equation}
which has five discrete eigenstates. The lowest order two are given by the symmetric ground state
$\eta_0(\xi)=\sech^5(\xi) , \lambda_0=-9 \alpha^2 $ and  the asymmetric first excited state $\eta_1(\xi)=\sech^4(\xi)\tanh(\xi) , \lambda_1=0 $. The latter is created by a simple shift of the original structure and is also known under Goldstone mode. Its is easily seen that our solution corresponds to the shift or Goldstone mode.

We can therefore conclude that a solitary electron hole (or bright soliton) exhibits marginal stability when derived from the S-equation. The undamped perturbation in this approach corresponds exactly to the first excited state, the shift mode, in Schamel's { kinetic theory of restricted transverse instability \cite{Schamel:1982}, where, by applying the ground state, instability was found to be present only transversely. The use of this shift mode $\eta_1$ instead of $\eta_0$ gives a new relevance to this analysis, although it is actually based on an unproven approximation, the truncation or fluid approach \cite{Lewis:1979}, in the generally still unsolved kinetic VP-linear stability problem \cite{Schamel:1982}. We mention that such an asymmetric shift mode was also recently observed in a PIC simulation \cite{Hutchinson:2018,Hutchinson:2019}. However, in contrast to the author, we see neither a justification nor a need for a new stability analysis dealing with the kinetic jetting of marginally passing electrons, since regularly passing particles are naturally taken into account in our analysis of the transverse instability.

 We note that our result is in agreement with the previous macroscopic marginal stability analyses of solitons by Kuznetsov \cite{Kuznetsov:1984} and of periodic structures by Bronski et al. \cite{Bronski:2011} in KdV-like equations.

Finally, we would like to point out, admittedly somewhat speculatively, the universal character of marginal stability of electrostatic structures in general. The reason for this is that marginal stability could also be seen microscopically, namely for single harmonic wave structures \cite{Schamel:2018,Schamel:2023}. However, to prove marginal stability for the entire spectrum of structures, including solitary and cnoidal waves, the solution of a non-local eigenvalue problem $\Big( (5),(6)  $ of $  \cite{Schamel:1982}; (18),(24) $ of $ \cite{Schamel:2018} $ or $ (55) $ of $ \cite{Schamel:2023}  \Big)$ is required, which is still pending except for harmonic structures, where the marginally stable perturbation can be considered as a shift mode (non-validity of Landau approach!).  

But maybe someone will come along who can prove the marginal stability for a general $\phi_0(x)$ including undisclosed potentials by using the shift mode in this eigenvalue problem as the eigenmode perturbation $\phi_1(x)=\phi_0'(x)$ and thus free the Schamel's restricted transversal instability theory from its limitation?  A big advantage would be that the x-dependence in all subsequent formulas could be replaced by a $\phi_0$-dependence. This holds for the non-local integral of (6) of \cite{Schamel:1982}, where $dx$ could be replaced by $\frac{d\phi_0}{\pm \sqrt{-2\mathcal V(\phi_0)}}$, but also for the $x$-derivative in (7) of  \cite{Schamel:1982} where $\partial_x$ could be replaced by $\pm\sqrt{-2\mathcal V(\phi_0)} \frac{d}{d\phi_0}$. This would therefore also open the door for undisclosed potentials $\phi_0(x)$, which represent by far the majority of potentials.

}

\section{Conclusion}
In summary, by focusing on one electron trapping scenario, we were able to re-derive the $S$-equation for the entire spectrum of cnoidal electron holes, including bright and dark solitons, encompassing structures near equilibria propagating at acoustic phase velocities. The single harmonic wave as a solution of the linearized $S$-equation did thereby not pose a problem with the required non-negativity, since it did not involve $\sin x$, $\cos x$, or other linear wave solutions such as Airy functions, but $(1+\cos x)/2$, which is non-negative. One important point is that all solutions are nonlinear solutions of the underlying phase space dynamics. This is easily seen in the harmonic SEA wave case where $B_e\approx (1 - \beta - v_0^2)=0$ requires an electron trapping parameter $\beta < -0.71$, which represents a depressed region around the phase velocity in $f_e(x,v)$. The parameter $\beta$ is thereby related to the first order perturbative trapping scenario (TS), see (1) of  \cite{Schamel:2023}. As a nonlinear solution of the VP system, this harmonic wave is not subject to Landau damping or Landau growth, for example in a current-carrying plasma, but is microscopically stable, independent of the strength of the current or the drift velocity \cite{Schamel:2023}.

However, the limits and extensions of the S-equation have not yet been thoroughly explored and certainly require further investigation. One example is the numerically observed, time-limited acceleration of a soliton during its propagation \cite{Mandal:2018}. It is known from the fundamental theory \cite{Schamel:2023} that a soliton in a higher velocity state can assume a lower energy, which may even become negative. Therefore, if it is possible to somehow incorporate energy conservation into the S-equation, there may be a chance to describe this time-limited acceleration effect within the framework of a modified version of the S-equation.

In \cite{Schamel:2020}, the focus is on an extension of the $S$-equation by an additional nonlinear term of logarithmic type, which is due to a further TS. Since there are a multitude of TSs, the $S$-equation is open for further nonlinear extensions. However, non-negativity remains a problem.

An important aspect of multiple simultaneously active TSs is that $\phi(x)$ becomes a mathematically unknown function, leading to an almost unlimited variety of structures; a point that has not yet been recognized as such by the plasma community.

To conclude, the $S$-equation as a macroscopic representation of deeper microscopic dynamics cannot, of course, reflect all aspects of phase space dynamics, so it is up to future generations to explore its scope. The coalescence of holes will clearly be outside their scope of application.

{
\appendix
\section{The Schamel method in a compact form}
Our main goal is to derive equations (4),(5) and thus to search for stationary solutions of the Vlasov-Poisson (VP) system using the method developed by Schamel in \cite{Schamel:1972}. This VP system consists of the Vlasov equation for electrons and ions as well as  Poisson's equation and is for normalized quantities in the rest frame of the wave given by
\begin{eqnarray}
[v\partial_x + \phi'(x)\partial_v] f_e(x,v)=0 \qquad
[u\partial_x -\theta \phi'(x) \partial_u] f_i(x,u)=0 \qquad \phi''(x)=\int \mathrm{d}v f_e(x,v) -\int \mathrm{d}u f_i(x,u)
\end{eqnarray}
A solution of the Vlasov equations is provided by the two sets of constants of motion: $\epsilon_e=\frac{v^2}{2} -\phi$, $ \sigma_e=\frac{v}{|v|}$ and $\epsilon_i=\frac{u^2}{2} -\theta(\psi-\phi)$, $ \sigma_i=\frac{u}{|u|}$,  respectively, in which the sign constants refer to untrapped particles only. It is given for an unperturbed Maxwellian plasma, assuming a positive pedestal $0\le \phi(x) \le \psi$ and the most smooth trapping scenario \cite{Schamel:2023}, by the Ansatz
\begin{eqnarray}
f_e(x,v)=\frac{1+k_0^2\psi}{\sqrt{2\pi}}\Big [\theta(\epsilon_e) \exp\Big (-\frac{1}{2}(\sigma_e \sqrt{2\epsilon_e}-
v_0)^2\Big ) +\theta(-\epsilon_e)\exp(-v_0^2/2)\exp(-\beta \epsilon_e)\Big ]
\end{eqnarray}
\begin{eqnarray}
f_i(x,u)=\frac{1+K_i}{\sqrt{2\pi}}\Big [\theta(\epsilon_i) \exp\Big (-\frac{1}{2}(\sigma_i \sqrt{2\epsilon_i}-
u_0)^2\Big ) +\theta(-\epsilon_i)\exp(-u_0^2/2)\exp(-\alpha \epsilon_i)\Big ]
\end{eqnarray}

The first part $\theta(\epsilon_{e,i})$ refers to free or passing particles, the second part $\theta(-\epsilon_{e,i})$ to
trapped particles. It holds $u_0=\sqrt{\frac{\theta}{\delta}} v_0$, and $v_0$ in $f_e(x,v)$ has to be replaced by $\tilde v_D:=|v_D-v_0|$ in case of a current-carrying plasma with a finite drift velocity $v_D$ between electrons and ions.
To obtain the densities, we have to integrate over the entire space. 

For small amplitudes, $\psi<<1$, we get
\begin{eqnarray}
n_e=1 + \frac{k_0^2}{2}\psi -\frac{1}{2}Z_r'(\frac{v_0}{\sqrt 2})\phi  - \frac{5 B_e}{4\sqrt\psi}\phi^{3/2} + ...
\end{eqnarray}

\begin{eqnarray}
n_i=1 + K_i  -\frac{\theta}{2}Z_r'(\frac{u_0}{\sqrt 2})(\psi-\phi)  - \frac{5 B_i}{4\sqrt\psi}[\theta(\psi-\phi)]^{3/2} + ...
\end{eqnarray}
In these equations we have defined:

\begin{eqnarray}
B_e:=\frac{16(1-\beta-v_0^2)}{15 \sqrt \pi}\exp(-v_0^2/2) \sqrt \psi  \qquad  B_i:=\frac{16(1-\alpha-u_0^2)}{15 \sqrt \pi}\exp(-u_0^2/2) \sqrt \psi
\end{eqnarray}
Both expressions coincide e.g. with (20b) of \cite{Schamel:2000}.
In the solitary wave limit, $k_0^2 \rightarrow 0$, both densities have to be equal at infinity, i.e. at $\phi=0$. This demand provides $K_i$ which becomes $K_i=\Big(\frac{1}{2}Z_r'(\frac{u_0}{\sqrt 2}) + \frac{5B_i\sqrt\theta}{4}\Big)\theta \psi$. After inserting $K_i$ into (26) the ion density simplifies to
\begin{eqnarray}
n_i=1 + \frac{\theta}{2}Z_r'(\frac{u_0}{\sqrt 2})\phi  + \frac{5 B_i \theta^{3/2}\psi}{4}[1-(1-\frac{\phi}{\psi})^{3/2}] + ...
\end{eqnarray}
To solve Poisson's equation, we introduce the (provisional) pseudo-potential in formal analogy to classical mechanics:
$\mathcal V_0(\phi;v_0)$ by $\phi''(x) = n_e(\phi) - n_i(\phi)=: -\mathcal V_0'(\phi;v_0)$, where the derivative refers to  $\phi$. We get
$ -\mathcal V_0'(\phi;v_0) =\frac{k_0^2}{2}\psi -[\frac{1}{2}Z_r'(\frac{v_0}{\sqrt 2}) +\frac{\theta}{2}Z_r'(\frac{u_0}{\sqrt 2})]\phi +\frac{5}{4\sqrt\psi}\Big( B_e \phi^{3/2} +B_i(\theta \psi)^{3/2}[1-(1-\frac{\phi}{\psi})^{3/2}] \Big) $. By $\phi$ integration, assuming that $\mathcal V_0(\phi;v_0)$ vanishes at $\phi=0$ we get 

\begin{eqnarray}
 -\mathcal V_0(\phi;v_0) =\frac{k_0^2}{2}\psi \phi  - [\frac{1}{2}Z_r'(\frac{v_0}{\sqrt 2}) +\frac{\theta}{2}Z_r'(\frac{u_0}{\sqrt 2})]\frac{\phi^2}{2} -\frac{B_e}{2\sqrt\psi} \phi^{5/2} - \frac{5B_i \theta^{3/2}\psi}{4} \Big(\phi-\frac{2\psi}{5} + \frac{2\psi}{5} (1-\frac{\phi}{\psi})^{5/2} \Big) 
\end{eqnarray}
 
By x-integration of the Poisson equation we obtain the pseudo-energy: $$ \qquad \frac{\phi'(x)^2}{2}+\mathcal V_0(\phi;v_0)=0.$$ Since at the potential maximum $\phi=\psi$ the slope of $\phi(x)$ (or the first derivative $\phi'(x)$) vanishes, we arrive directly at: $\mathcal V_0(\psi;v_0)=0$,  which is a determining equation for $v_0$.

This equation is commonly referred to as the nonlinear dispersion relation (NDR) and is:
\begin{eqnarray}  \label{eq30}
k_0^2 - \frac{1}{2}Z_r'(\frac{v_0}{\sqrt 2}) -\frac{\theta}{2}Z_r'(\frac{u_0}{\sqrt 2}) = B_e + \frac{3}{2} B_i \theta^{3/2}
\end{eqnarray}
Its solution $v_0$ provides the first part of our problem of finding a suitable $\phi(x-v_0t)$. The second part, the determination of the shape of $\phi(x)$, follows directly from the canonical form of the pseudo-energy:
\begin{eqnarray}
  \qquad \frac{\phi'(x)^2}{2}+\mathcal V(\phi)=0
\end{eqnarray}

The canonical pseudo-potential $\mathcal V(\phi)$ is thereby obtained by replacing the $v_0$-dependent part in $\mathcal V_0(\phi;v_0)$ by the NDR and is:
\begin{eqnarray} \label{eq32}
 -\mathcal V(\phi) =\frac{k_0^2}{2} \phi ( \psi-\phi)   +\frac{B_e}{2\sqrt\psi}\phi^2(\sqrt \psi- \sqrt\phi) + \frac{B_i \theta^{3/2}}{4\sqrt \psi} \Big(\phi\sqrt\psi(3\phi-5\psi) +2[\psi^{5/2} -(\psi-(\psi-\phi)^{5/2}] \Big) 
\end{eqnarray}
Equations (\ref{eq30})-(\ref{eq32}) provide the general solution of our problem with two trapping scenarios $B_e,B_i$. The NDR and the pseudo-potential are identical to earlier expressions such as (24),(25) or (44),(45) of \cite{Schamel:2000} or (51),(52) of \cite{Schamel:2023}, respectively. 

In the simplified case without ion trapping effect, $B_i=0$, we therefore obtain our desired result (\ref{eq1}),(\ref{eq2}). To avoid misunderstandings: $B_i=0$ does not mean the absence of ion trapping, but the absence of their effects, since it holds $\alpha=1-u_0^2$ and thus ion trapping still participates.


 }

\end{document}